\documentclass[11pt]{article}
\usepackage{fullpage,graphicx}

\newtheorem{theorem}{Theorem}
\newtheorem{lemma}{Lemma}

\newcommand{\Oh}[1]
    {\ensuremath{\mathcal{O}\!\left( {#1} \right)}}

\begin{document}

\title{Heaviest Induced Ancestors and\\Longest Common Substrings}
\author{Travis Gagie\thanks{University of Helsinki and HIIT}
    \and Pawe\l\ Gawrychowski\thanks{Max-Planck-Institut f\"{u}r Informatik}
    \and Yakov Nekrich\thanks{University of Kansas}}
\index{Gagie, Travis}
\index{Gawrychowski, Pawe\l}
\index{Nekrich, Yakov}
\thispagestyle{empty}
\maketitle

\begin{abstract}
Suppose we have two trees on the same set of leaves, in which nodes are weighted such that children are heavier than their parents.  We say a node from the first tree and a node from the second tree are induced together if they have a common leaf descendant.  In this paper we describe  data structures that efficiently support the following heaviest-induced-ancestor query: given a node from the first tree and a node from the second tree, find an induced pair of their ancestors with maximum combined weight.  Our solutions are based on a geometric interpretation that enables us to find heaviest induced ancestors using range queries.  We then show how to use these results to build an LZ-compressed index with which we can quickly find with high probability a longest substring common to the indexed string and a given pattern.
\end{abstract}

\section{Introduction} \label{sec:introduction}

In their paper ``Range Searching over Tree Cross Products'', Buchsbaum, Goodrich and Westbrook~\cite{BGW00} considered how, given a forest of trees \(T_1, \ldots, T_d\) and a subset $E$ of the cross product of the trees' node sets, we can preprocess the trees such that later, given a $d$-tuple $u$ consisting of one node from each tree, we can, e.g., quickly determine whether there is any $d$-tuple \(e \in E\) that {\em induces} $u$ --- i.e., such that every node in $e$ is a descendant of the corresponding node in $u$.  (Unfortunately, some of their work was later found to be faulty; see~\cite{ALLS07}.)

In this paper we assume we have two trees $T_1$ and $T_2$ on the same set of $n$ leaves, in which each internal node has at least two children and nodes are weighted such that children are heavier than their parents.  We assume $E$ is the identity relation on the leaves.  Following Buchsbaum et al., we say a node in $T_1$ and a node in $T_2$ are {\em induced} together if they have a common leaf descendant.  We consider how, given a node $v_1$ in $T_1$ and a node $v_2$ in $T_2$, we can quickly find a pair of their {\em heaviest induced ancestors} (HIAs) --- i.e., an ancestor $u_1$ of $v_1$ and ancestors $u_2$ of $v_2$ such that $u_1$ and $u_2$ are induced together and have maximum combined weight.

\begin{figure*}[t]
\begin{center}
\resizebox{.8\textwidth}{!}
{\begin{tabular}{c@{\hspace{10ex}}c}
\includegraphics[width = 50ex]{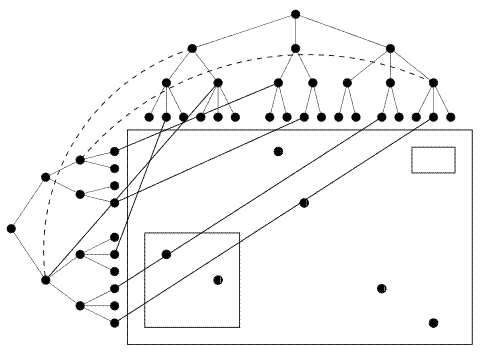}&
\includegraphics[width = 50ex]{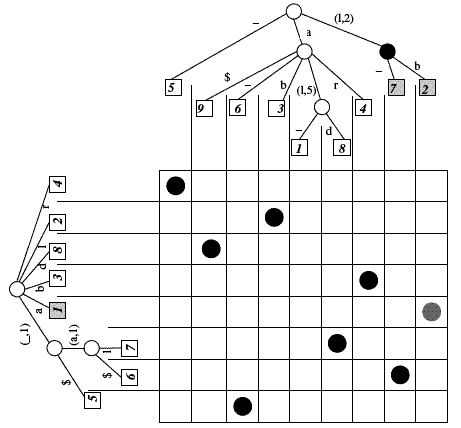}
\end{tabular}}
\caption{Figure 1 from Buchsbaum et al.'s ``Range Searching over Tree Cross Products'' and Figure 2b from Kreft and Navarro's ``On Compressing and Indexing Repetitive Sequences'', whose similarity suggests a link between the two problems.  We exploit this link when we use HIA queries to implement LCS queries.}
\label{fig:comparison}
\end{center}
\end{figure*}

In Section~\ref{sec:HIAs} we give several tradeoffs for data structures supporting HIA queries: e.g., we describe an $\Oh{n}$-space data structure with $\Oh{\log^3 n (\log \log n)^2}$ query time.  Our motivation is the problem of building LZ-compressed indexes with which we can quickly find a longest common substring (LCS) of the indexed string and a given pattern.  Tree cross products and LZ-indexes may seem unrelated, until we compare figures from Buchbaum et al.'s paper and Kreft and Navarro's ``On Compressing and Indexing Repetitive Sequences'', shown in Figure~\ref{fig:comparison}.  In Section~\ref{sec:LCSs} we show how, given a string $S$ of length $N$ whose LZ77 parse~\cite{ZL77} consists of $n$ phrases, we can build an $\Oh{n \log N}$-space index with which, given a pattern $P$ of length $m$, we can find with high probability an LCS of $P$ and $S$ in $\Oh{m \log^2 n}$ time.

\section{Heaviest Induced Ancestors} \label{sec:HIAs}

An obvious way to support HIA queries is to impose orderings on $T_1$ and $T_2$; for each node $u$, store $u$'s weight and the numbers of leaves to the left of $u$'s leftmost and rightmost leaf descendants; and store a range-emptiness data structure for the \(n \times n\) grid on which there is a marker at point \((x, y)\) if the $x$-th leaf from the left in $T_1$ is the $y$-th leaf from the left in $T_2$.  Suppose there are \(x_1 - 1\) and \(x_2 - 1\) leaves to the left of the leftmost and rightmost leaf descendants of $u_1$ in $T_1$, and  \(y_1 - 1\) and \(y_2 - 1\) leaves to the left of the leftmost and rightmost leaf descendants of $u_2$ in $T_2$.  Then $u_1$ and $u_2$ are induced together if and only if the range \([x_1..x_2] \times [y_1..y_2]\) is non-empty.  Chan, Larsen and P\v{a}tra\c{s}cu~\cite{CLP11} showed how we can store the range-emptiness data structure in $\Oh{n}$ space with $\Oh{\log^\epsilon n}$ query time, or in $\Oh{n \log \log n}$ space with $\Oh{\log \log n}$ query time.

Given a node $v_1$ in $T_1$ and $v_2$ in $T_2$, we start with a pointer $p$ to $v_1$ and a pointer $q$ to the root of $T_2$.  If the nodes $u_1$ and $u_2$ indicated by $p$ and $q$ are induced together, then we check whether $u_1$ and $u_2$ have greater combined weight than any induced pair we have seen so far and move $q$ down one level toward $v_2$; otherwise, we move $p$ up one level toward the root of $T_1$; we stop when $p$ reaches the root of $T_1$ or $q$ reaches $v_2$.  This takes a total of $\Oh{\mathrm{depth} (v_1) + \mathrm{depth} (v_2)}$ range-emptiness queries.

\subsection{An $\Oh{n \log^2 n}$-space data structure with $\Oh{\log n \log \log n}$ query time} \label{subsec:big HIAs}

We now describe an $\Oh{n \log^2 n}$-space data structure with $\Oh{\log n \log \log n}$ query time; later we will show how to reduce the space via sampling, at the cost of increasing the query time.  We first compute the heavy-path decompositions~\cite{ST81} of $T_1$ and $T_2$.  These decompositions have the property that every root-to-leaf path consists of the prefixes of $\Oh{\log n}$ heavy paths.  Therefore, for each leaf $w$ there are $\Oh{\log^2 n}$ pairs \((a, b)\) such that $a$ and $b$ are the lowest nodes in their heavy paths in $T_1$ and $T_2$, respectively, that are ancestors of $w$.

For each pair of heavy paths, one in $T_1$ and the other in $T_2$, we store a list containing each pair \((a, b)\) such that $a$ is a node in the first path, $b$ is a node in the second path, $a$ and $b$ are induced together by some leaf, $a$'s child in the first path is not induced with $b$ by any leaf, and $b$'s child in the second path is not induced with $a$ by any leaf.  We call this the paths' {\em skyline list}.  Since there are $n$ leaves and $\Oh{\log^2 n}$ pairs for each leaf, all the skyline lists have total length $\Oh{n \log^2 n}$.  We store a perfect hash table containing the non-empty lists.

Let \(L = (a_1, b_1), \ldots, (a_\ell, b_\ell)\) be the skyline list for a pair of heavy paths, sorted such that \(\mathrm{depth} (a_1) > \cdots > \mathrm{depth} (a_\ell)\) and \(\mathrm{weight} (a_1) > \cdots > \mathrm{weight} (a_\ell)\) or, equivalently, \(\mathrm{depth} (b_1) < \cdots < \mathrm{depth} (b_\ell)\) and \(\mathrm{weight} (b_1) < \cdots < \mathrm{weight} (b_\ell)\).  (Notice that, if $a$ is induced with $b$, then every ancestor of $a$ is also induced with $b$.  Therefore, if \((a_i, b_i)\) and \((a_j, b_j)\) are both pairs in $L$ and $a_i$ is deeper than $a_j$ then, by our definition of a pair in a skyline list, $b_j$ must be deeper than $b_i$.)  Let $v_1$ be a node in the first path and $v_2$ be a node in the second path.  Suppose we want to find the pair of induced ancestors in these paths of $v_1$ and $v_2$ with maximum combined weight.  With the approach described above, we would start with a pointer $p$ to $v_1$ and a pointer $q$ to the highest node in the second path, then move $p$ up
toward the highest node in the first path and $q$ down toward $v_2$.

\begin{figure*}[t]
\begin{center}
\resizebox{.6\textwidth}{!}
{\begin{tabular}{c@{\hspace{8ex}}c@{\hspace{8ex}}c}
\includegraphics[width = 30ex]{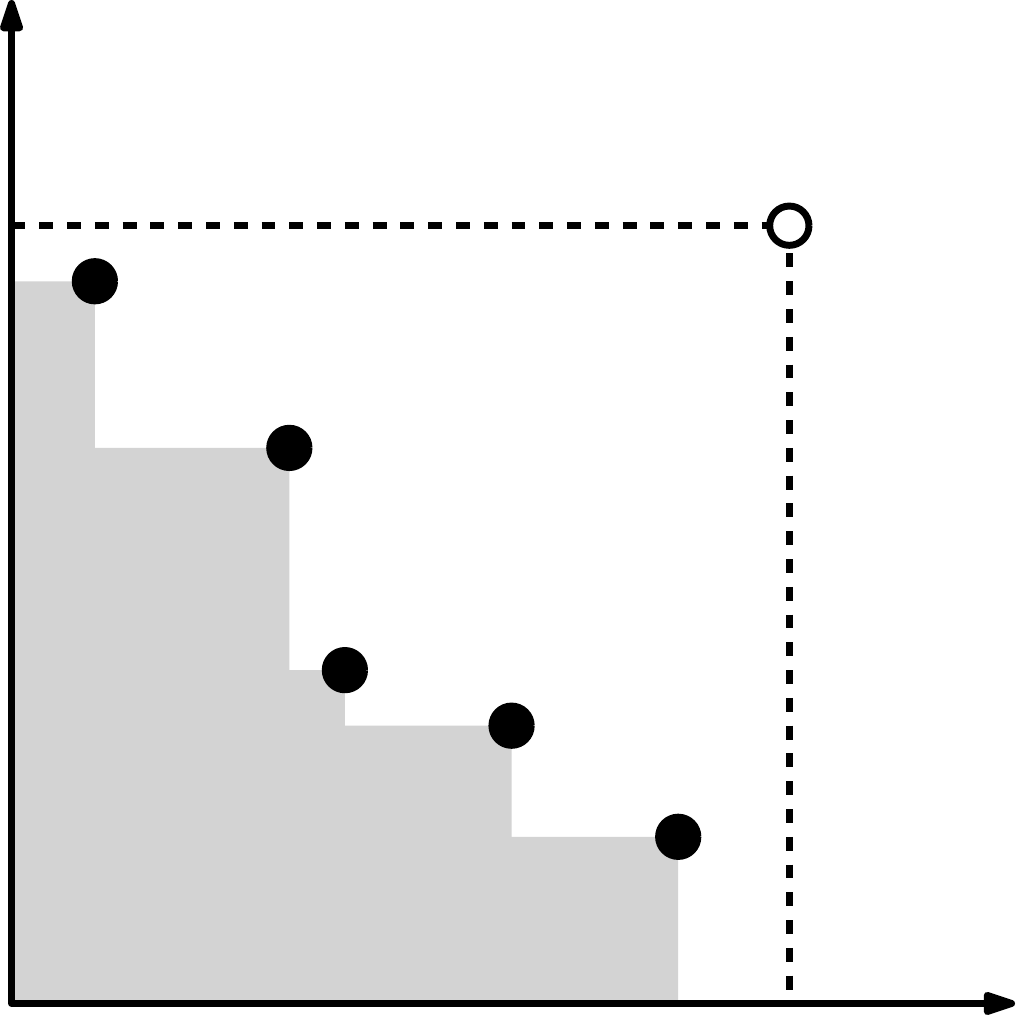}&
\includegraphics[width = 30ex]{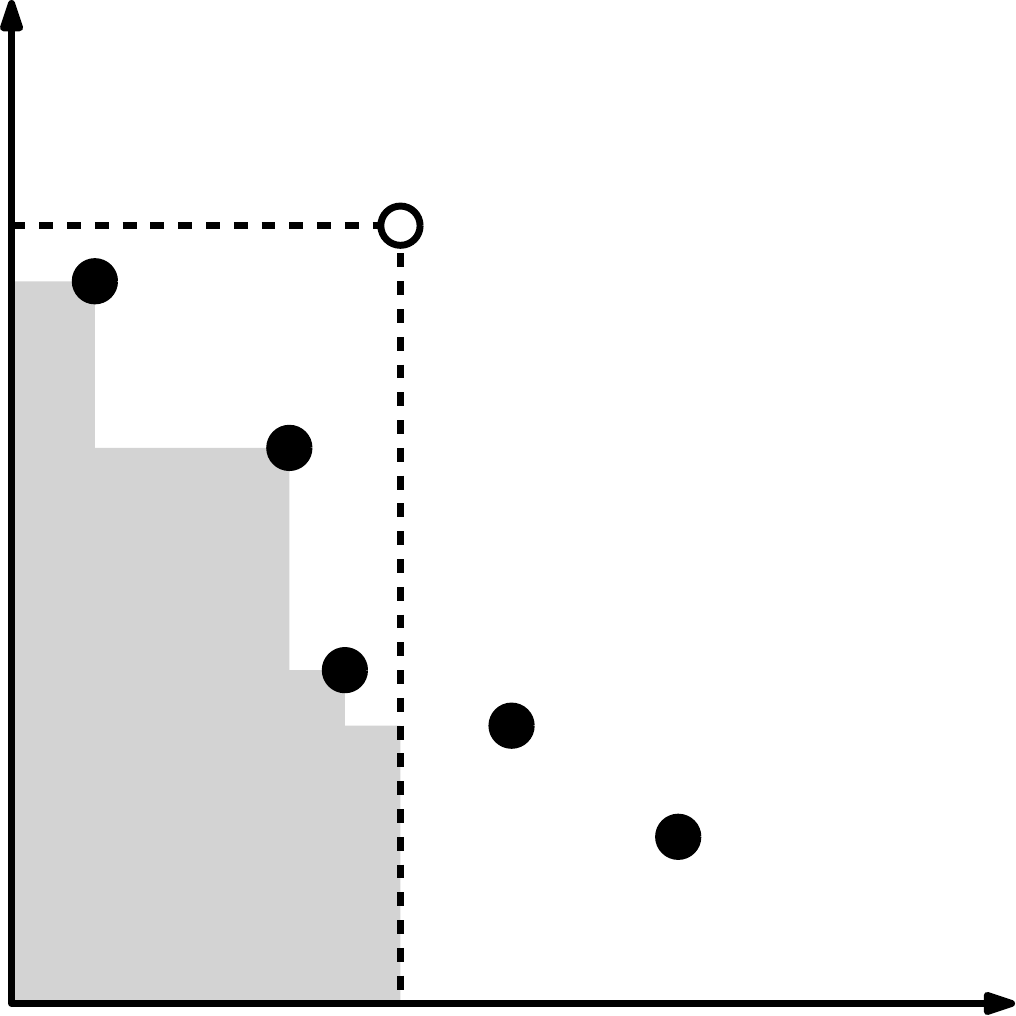}&
\includegraphics[width = 30ex]{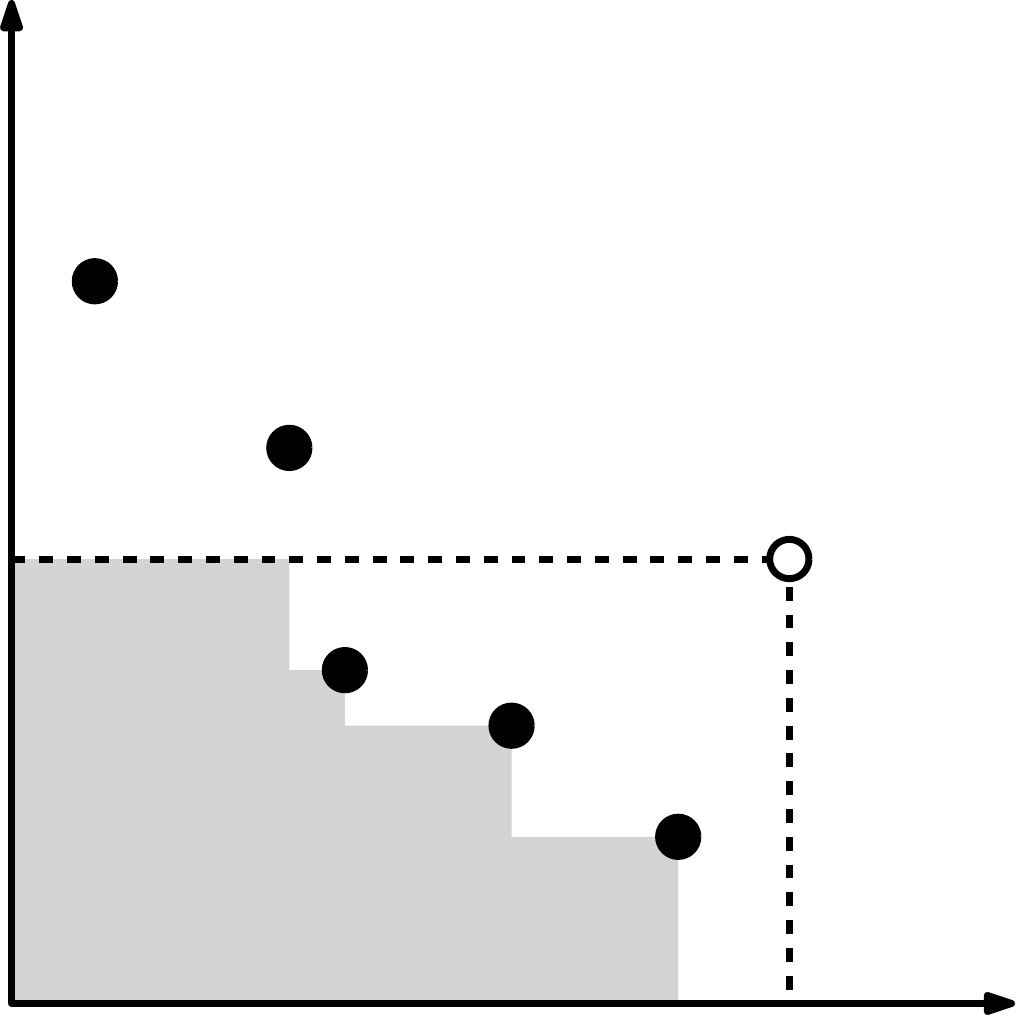}
\end{tabular}}

\vspace{5ex}

\resizebox{.6\textwidth}{!}
{\begin{tabular}{@{\hspace{19ex}}c@{\hspace{8ex}}c@{\hspace{19ex}}}
\includegraphics[width = 30ex]{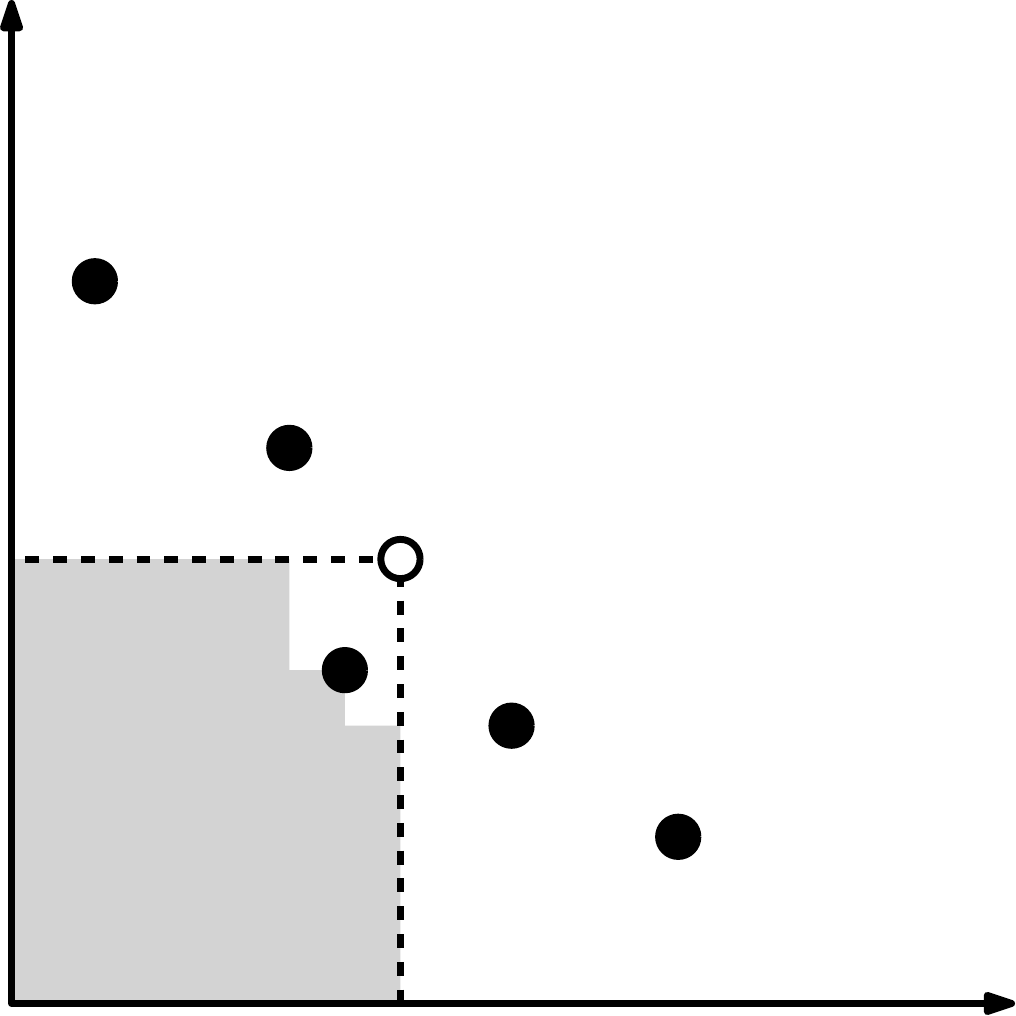}&
\includegraphics[width = 30ex]{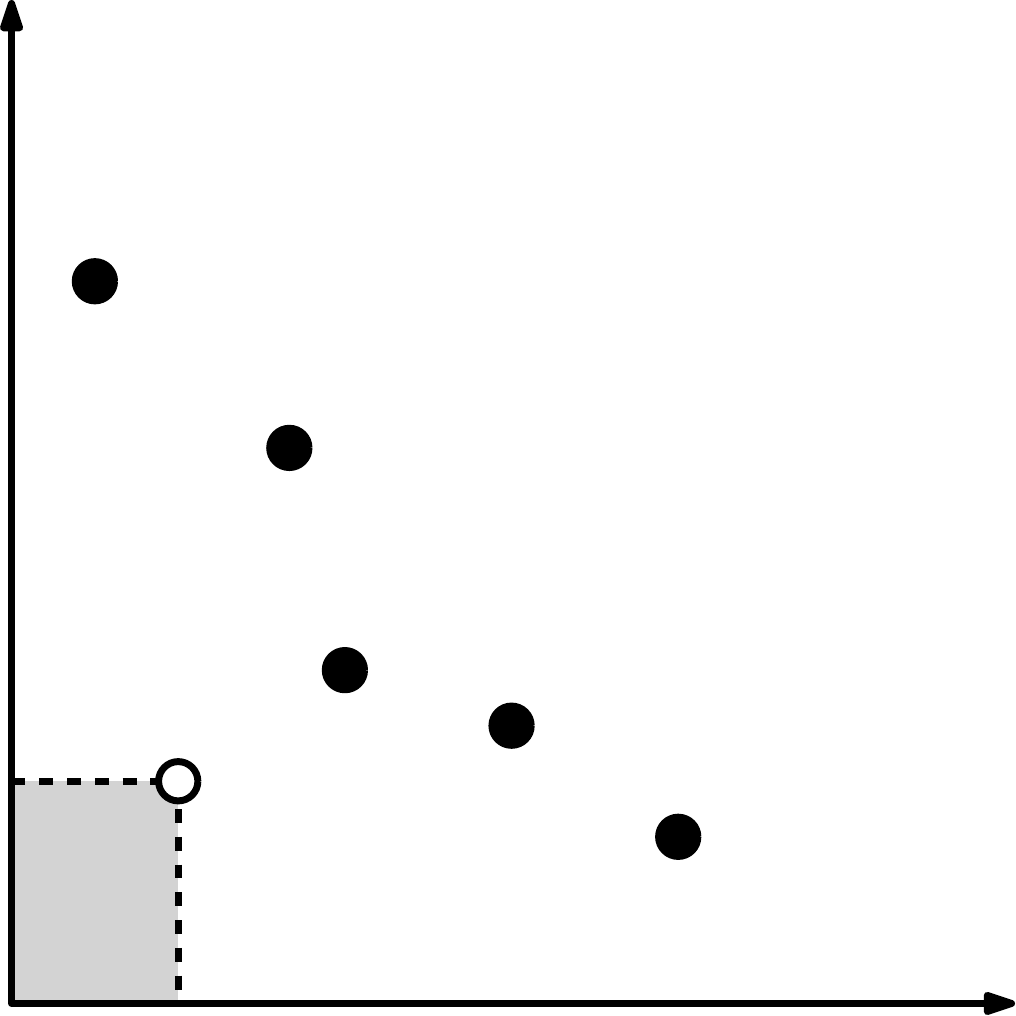}
\end{tabular}}
\caption{Finding the pair of induced ancestors of $v_1$ and $v_2$ with maximum combined weight is equivalent to storing a skyline such that, given a query point, we can quickly find the point \((x, y)\) dominated both by some point on the skyline and by the query point, such that \(x + y\) is maximized.}
\label{fig:grid}
\end{center}
\end{figure*}

A geometric visualization is shown in Figure~\ref{fig:grid}: the filled markers (from right to left) have coordinates \((\mathrm{weight} (a_1), \mathrm{weight} (b_1)), \ldots, (\mathrm{weight} (a_\ell), \mathrm{weight} (b_\ell))\), the hollow marker has coordinates \((\mathrm{weight} (v_1), \mathrm{weight} (v_2))\), and we seek the point \((x, y)\) that is dominated both by some filled marker and by the hollow marker, such that \(x + y\) is maximized.  Notice
\[(\mathrm{weight} (a_1), \mathrm{weight} (b_1)), \ldots, (\mathrm{weight} (a_\ell), \mathrm{weight} (b_\ell))\]
is a skyline --- i.e., no marker dominates any other marker.  There are five cases to consider: neither $v_1$ nor $v_2$ are induced with any other nodes in the paths; $v_1$ is induced with some node in the second path, but $v_2$ is not induced with any node in the first path; $v_1$ is not induced with any node in the second path, but $v_2$ is induced with some node in the first path; both $v_1$ and $v_2$ are induced with some nodes in the
paths, but not with each other; and $v_1$ and $v_2$ are induced together.

It follows that finding the pair of induced ancestors in these paths of $v_1$ and $v_2$ with maximum combined weight is equivalent to finding the interval \((a_i, b_i), \ldots, (a_j, b_j)\) in $L$ such that \(\mathrm{depth} (a_{i - 1}) > \mathrm{depth} (v_1) \geq \mathrm{depth} (a_i)\) and \(\mathrm{depth} (b_j) \leq \mathrm{depth} (v_2) < \mathrm{depth} (b_{j + 1})\), then finding the maximum in
\[\begin{array}{c c c}
\mathrm{weight} (v_1) &+& \mathrm{weight} (b_{i - 1}),\\
\mathrm{weight} (a_i) &+& \mathrm{weight} (b_i),\\
\mathrm{weight} (a_{i+1}) &+& \mathrm{weight} (b_{i+1}),\\
 &\vdots& \\
\mathrm{weight} (a_{j-1}) &+& \mathrm{weight} (b_{j-1}),\\
\mathrm{weight} (a_j) &+& \mathrm{weight} (b_j),\\
\mathrm{weight} (a_{j + 1}) &+& \mathrm{weight} (v_2)\,.
\end{array}\]

Therefore, if we store $\Oh{\ell}$-space predecessor data structures with $\Oh{\log \log n}$ query time~\cite{BKZ77} for \(\mathrm{depth} (a_1), \ldots, \mathrm{depth} (a_\ell)\) and \(\mathrm{depth} (b_1), \ldots, \mathrm{depth} (b_\ell)\) and an $\Oh{\ell}$-space range-maximum data with $\Oh{1}$ query time~\cite{FH11} for \(\mathrm{weight} (a_1) + \mathrm{weight} (b_1), \ldots, \mathrm{weight} (a_\ell) + \mathrm{weight} (b_\ell)\), then in $\Oh{\log \log n}$ time we can find the pair of induced ancestors in these paths of $v_1$ and $v_2$ with maximum combined weight.  Notice that we can assign $v_1$ and $v_2$ different weights when finding this pair of induced ancestors; this will be useful in Section~\ref{sec:LCSs}.

\begin{lemma} \label{lem:intra-list}
We can store $T_1$ and $T_2$ in $\Oh{n \log^2 n}$ space such that, given nodes $v_1$ in $T_1$ and $v_2$ in $T_2$, in $\Oh{\log \log n}$ time we can find a pair of their induced ancestors in the same heavy paths with maximum combined weight, if such a pair exists.
\end{lemma}

To find a pair of HIAs of $v_1$ and $v_2$, we consider the heavy-path decompositions of $T_1$ and $T_2$ as trees $\mathbf{T}_1$ and $\mathbf{T}_2$ of height $\Oh{\log n}$ in which each node is a heavy path and $V$ is a child of $U$ in $\mathbf{T}_1$ or $\mathbf{T}_2$ if the highest node in the path $V$ is a child of a node in the path $U$ in $T_1$ or $T_2$.  We start with a pointer $\mathbf{p}$ to the path $V_1$ containing $v_1$ and a pointer $\mathbf{q}$ to the root of $\mathbf{T}_2$.  If the skyline list for the nodes $U_1$ and $U_2$ indicated by $\mathbf{p}$ and $\mathbf{q}$ is non-empty, then we apply Lemma~\ref{lem:intra-list} to the deepest ancestors of $v_1$ and $v_2$ in $U_1$ and $U_2$, check whether the induced ancestors we find have greater combined weight than any induced pair we have seen so far and move $\mathbf{q}$ down one level toward $V_2$ (to execute the descent efficiently, in the very beginning we generate the whole path from $V_2$ containing $v_2$ to the root of $\mathbf{T}_2$);
otherwise, we move $\mathbf{p}$ up one level toward the root of $\mathbf{T}_1$.  This takes a total of $\Oh{\log n \log \log n}$ time.  Again, we have the option of assigning $v_1$ and $v_2$ different weights for the purpose of the query.

\begin{theorem} \label{thm:big HIAs}
We can store $T_1$ and $T_2$ in $\Oh{n \log^2 n}$ space such that, given nodes $v_1$ in $T_1$ and $v_2$ in $T_2$, in $\Oh{\log n \log \log n}$ time we can find a pair of their HIAs.
\end{theorem}

In the full version of this paper we will reduce the query time in Theorem~\ref{thm:big HIAs} to $\Oh{\log n}$ via fractional cascading~\cite{CG86}; however, this is not straightforward, as we need to modify our approach such that predecessor searches keep the same target as we change pairs of heavy paths and the hive or catalogue graph has bounded degree.

\subsection{An $\Oh{n \log n}$-space data structure with $\Oh{\log^2 n}$ query time} \label{subsec:small HIAs}

To reduce the space bound in Theorem~\ref{thm:big HIAs} to $\Oh{n \log n}$, we choose the orderings to impose on $T_1$ and $T_2$ such that each heavy path consists either entirely of leftmost children or entirely of rightmost children (except possibly for the highest nodes).  We store an $\Oh{n \log^\epsilon n}$-space data structure~\cite{ABR00} that supports $\Oh{\log \log n + k}$-time range-reporting queries on the grid described at the beginning of this section, where $k$ is the number of points reported.

\begin{figure}[t]
\begin{center}
\includegraphics[width = 30ex]{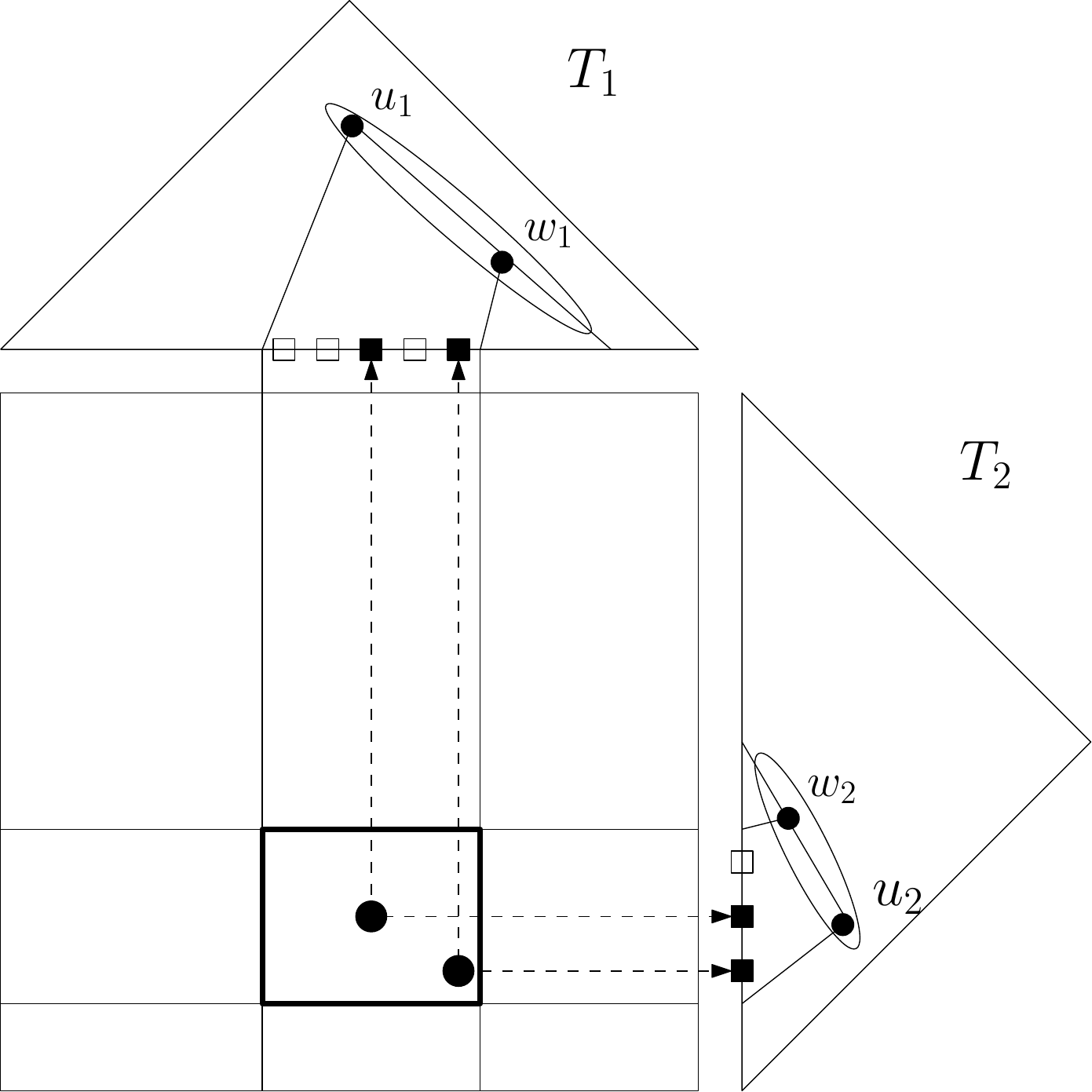}
\caption{Suppose $u_1$ is an ancestor of $w_1$ in the same heavy path (shown as an oval) in $T_1$ and $u_2$ is an ancestor of $w_2$ in the same heavy path (also shown as an oval) in $T_2$.  We can use a range-reporting query to find the leaves (shown as filled boxes) that induce $u_1$ and $u_2$ together but not $w_1$ and $w_2$ together.}
\label{fig:reporting}
\end{center}
\end{figure}

Notice that, if $u_1$ is an ancestor of $w_1$ in the same heavy path in $T_1$ and $u_2$ is an ancestor of $w_2$ in the same heavy path in $T_2$, then we can use a range-reporting query to find, e.g., the leaves that induce $u_1$ and $u_2$ together but not $w_1$ and $w_2$ together.  Suppose there are \(x_1 - 1\) and \(x_2 - 1 > x_1 - 1\) leaves to the left of the leftmost leaf descendants of $u_1$ and $w_1$ in $T_1$, and  \(y_1 - 1\) and \(y_2 - 1 > y_2 - 1\) leaves to the left of the rightmost leaf descendants of $w_2$ and $u_2$ in $T_2$; the cases when \(x_2 < x_1\) or \(y_2 < y_1\) are symmetric.  Then the leaves that induce $u_1$ and $u_2$ together but not $w_1$ and $w_2$ together are indicated by markers in \([x_1..x_2 - 1] \times [y_1..y_2 - 1]\), as illustrated in Figure~\ref{fig:reporting}.  That is, we query the cross product of the ranges of leaves in the subtrees of $u_1$ and $u_2$ but not $w_1$ and $w_2$.  Similarly, we can find the leaves that induce $u_1$ and $w_2$ together but not $u_2$ and $w_1$ together (or vice versa), but then we query the cross product of the ranges of leaves in the subtrees of $u_1$ and $w_2$ but not $w_1$ (or of $u_2$ and $w_1$ but not $w_2$).

For each pair of heavy paths, we build a list containing each pair \((a, b)\) such that, for some leaf $x$, $a$ is the lowest ancestor of $x$ in the first path and $b$ is the lowest ancestor of $x$ in the second path.  We call this the paths' {\em extended list}, and consider it in decreasing order by the depth of the first component.  Notice that an extended list is a supersequence of the the corresponding skyline list, but all the extended lists together still have total length $\Oh{n \log^2 n}$.

We do not store the complete extended lists; instead, we sample only every \((\log n)\)-th pair, so the sampled lists take $\Oh{n \log n}$ space.  We store a perfect hash function containing the non-empty sampled lists; we can still tell if a list was empty before sampling by using a range-reporting query to find any common leaf descendants of the highest nodes in the heavy paths.  Given two consecutive sampled pairs from an extended list, in $\Oh{\log n}$ time we can recover the pairs between them using a range-reporting query, as described above.

With each sampled pair from an extended list, we store the preceding and succeeding pairs (possibly unsampled) that also belong to the corresponding skyline list; recall that the extended list is a supersequence of the skyline list.  This gives us an irregular sampling (which may include duplicates) of pairs from the skyline lists, which has total size $\Oh{n \log n}$.  Instead of storing predecessor and range-maximum data structures over the complete skyline lists, we store them over these sampled skyline lists, so we use a total of $\Oh{n \log n}$ space.  Since these data structures are over sampled skyline lists, querying them indicates only which \((\log n)\)-length block in a complete extended list contain the pair that would be returned by a query on a corresponding complete skyline list.  We can recover any \((\log n)\)-length block of a complete extended list in $\Oh{\log n}$ time with a range-reporting query, however, and then scan that block to find the pair with maximum combined weight.

If we sample only every \((\log^2 n)\)-th pair from each extended list and use Chan et al.'s linear-space data structure for range reporting, then we obtain an even smaller (albeit slower) data structure for HIA queries.

\begin{theorem} \label{thm:small HIAs}
We can store $T_1$ and $T_2$ in $\Oh{n \log n}$ space such that, given nodes $v_1$ in $T_1$ and $v_2$ in $T_2$, in $\Oh{\log^2 n}$ time we can find a pair of their HIAs.  Alternatively, we can store $T_1$ and $T_2$ in $\Oh{n}$ space such that, given $v_1$ and $v_2$, in $\Oh{\log^{3 + \epsilon} n}$ time we can find a pair of their HIAs.
\end{theorem}

\section{Longest Common Substrings} \label{sec:LCSs}

LZ-compressed indexes can use much less space than compressed suffix arrays or FM-indexes (see~\cite{ANS12,KN13,MNSV10,NM07}) when the indexed string is highly repetitive (e.g., versioned text documents, software repositories or databases of genomes of individuals from the same species).  Although there is an extensive literature on the LCS problem, including Weiner's classic paper~\cite{Wei73} on suffix trees and more recent algorithms for inputs compressed with the Burrows-Wheeler Transform (see~\cite{OGK10}) or grammars (see~\cite{MIIS+09}), we do not know of any grammar- or LZ-compressed indexes designed to support fast LCS queries.

Most LZ-compressed indexes are based on an idea by K\"arkk\"ainen and Ukkonen~\cite{KU96}: we store a data structure supporting access to the indexed string \(S [1..N]\); we store one Patricia tree~\cite{Mor68} $T_\mathrm{rev}$ for the reversed phrases in the LZ parse, and another $T_\mathrm{suf}$ for the suffixes starting at phrase boundaries; we store a data structure for 4-sided range reporting for the grid on which there is a marker at point \((x, y)\) if the $x$-th phrase in right-to-left lexicographic order is followed in the parse by the lexicographically $y$-th suffix starting at a phrase boundary; and we store a data structure for 2-sided range reporting for the grid on which there is a marker at point \((x, y)\) if a phrase source begins at position $x$ and ends at position $y$.

Given a pattern \(P [1..m]\), for \(1 \leq i \leq m\) we search for \((P [1..i])^\mathrm{rev}\) in $T_\mathrm{rev}$ (where the superscript $\mathrm{rev}$ indicates that a string is reversed) and for \(P [i + 1..m]\) in $T_\mathrm{suf}$; access $S$ to check that the path labels of the nodes where the searches terminate really are prefixed by \((P [1..i])^\mathrm{rev}\) and \(P [i + 1..m]\); find the ranges of leaves that are descendants of those nodes; and perform a 4-sided range-reporting query on the cross product of those ranges.  This gives us the locations of occurrences of $P$ in $S$ that touch phrase boundaries.  We then use recursive 2-sided range-reporting queries to find the phrase sources covering the occurrences we have found so far.

Rytter~\cite{Ryt03} showed how, if the LZ77 parse of $S$ consists of $n$ phrases, then we can build a balanced straight-line program (BSLP) for $S$ with $\Oh{n \log N}$ rules.  A BSLP for $S$ is a context-free grammar in Chomsky normal form that generates $S$ and only $S$ such that, in the parse tree of $S$, every node's height is logarithmic in the size of its subtree.  We showed in a previous paper~\cite{GGKNP12a,GGKNP12b} how we can store a BSLP for $S$ in $\Oh{n \log N}$ space such that extracting a substring of length $m$ from around a phrase boundary takes $\Oh{m}$ time.  Using this data structure for access to $S$ and choosing the rest of the data structures appropriately, we can store $S$ in $\Oh{n \log N}$ space such that listing all the $\mathrm{occ}$ occurrences of $P$ in $S$ takes $\Oh{m^2 + \mathrm{occ} \log \log N}$ time.

Our solution can easily be modified to find the LCS of $P$ and $S$ in $\Oh{m^2 \log \log n}$ time: we store the BSLP for $S$; the two Patricia trees $T_\mathrm{rev}$ and $T_\mathrm{suf}$, with the nodes weighted by the lengths of their path labels; and an instance of Chan et al.'s $\Oh{n \log \log n}$-space range-emptiness data structure with $\Oh{\log \log n}$ query time, instead of the data structure for 4-sided range range reporting.  All these data structures together take a total of $\Oh{n \log N}$ space.  By the definition of the LZ77 parse, the first occurrence of every substring in $S$ touches a phrase boundary.  It follows that we can find the LCS of $P$ and $S$ by finding, for \(1 \leq i \leq m\), values $h$ and $j$ such that some phrase ends with \(P [h..i]\) and the next phrase starts with \(P [i + 1..j]\) and \(j - h + 1\) is maximum.

For \(1 \leq i \leq m\) we search for \((P [1..i])^\mathrm{rev}\) in $T_\mathrm{rev}$ and for \(P [i + 1..m]\) in $T_\mathrm{suf}$, as before; access $S$ to find the longest common prefix (LCP) of \((P [1..i])^\mathrm{rev}\) and the path label of the node where the search in $T_\mathrm{rev}$ terminates, and the LCP of \(P [i + 1..m]\) and the path label of the node where the search in $T_\mathrm{suf}$ terminates; take $v_1$ and $v_2$ to be the loci of those LCPs, and treat them as having weights equal to the lengths of the LCPs; and then use the range-emptiness data structure and the simple HIA algorithm described at the beginning of Section~\ref{sec:HIAs} to find $h$ and $j$ for this choice of $i$.  For each choice of $i$ this takes $\Oh{m \log \log n}$ time, so we use $\Oh{m^2 \log \log n}$ time in total.

\begin{lemma} \label{lem:slow LCSs}
We can store $S$ in $\Oh{n \log N}$ space such that, given a pattern $P$ of length $m$, we can find the LCS of $P$ and $S$ in $\Oh{m^2 \log \log n}$ time.
\end{lemma}

We now show how to use our data structure for HIA queries to reduce the dependence on $m$ in Lemma~\ref{lem:slow LCSs} from quadratic to linear.

Ferragina~\cite{Fer13} showed how, by storing path labels' Karp-Rabin hashes~\cite{KR87} and rebalancing the Patricia trees via centroid decompositions, in a total of $\Oh{m \log n}$ time we can find with high probability the nodes where the searches for \((P [1..i])^\mathrm{rev}\) and \(P [i + 1..m]\) terminate, for all choices of $i$.  In our previous paper we showed how, by storing the hash of the expansion of each non-terminal in the BSLP for $S$, in $\Oh{m \log m}$ time we can then verify with high probability that the path labels of the nodes where the searches terminate really are prefixed by \((P [1..i])^\mathrm{rev}\) and \(P [i + 1..m]\).

Using the same techniques, in $\Oh{m \log m}$ time we can find with high probability for all choices of $i$, the LCP of \((P [1..i])^\mathrm{rev}\) and any reversed phrase, and the LCP of \(P [i + 1..m]\) and any suffix starting at a phrase boundary.  If \(m = n^{\mathcal{O} (1)}\) then \(\Oh{m \log m} = \Oh{m \log n}\).  If \(m = n^{\omega (1)}\), then we can preprocess $P$ and batch the searches for the LCPs, to perform them all in $\Oh{m}$ time.

More specifically, to find the LCPs of \(P [2..m], \ldots, P [m..m]\) and suffixes starting at phrase boundaries, we first build the suffix array and LCP array of $P$.  For \(1 \leq i \leq m\) we use Ferragina's data structure to find the suffix starting at a phrase boundary whose LCP with \(P [i + 1..m]\) is maximum.  For each phrase boundary, we use the suffix array and LCP array of $P$ to build a Patricia tree for the suffixes of $P$ whose LCPs we will seek at that phrase boundary.  We then balance these Patricia trees via centroid decompositions.  For each phrase boundary, we determine the length of the LCP of any suffix of $P$ and the suffix starting at that phrase boundary. We then use the LCP array of $P$ to find the LCPs of \(P [2..m], \ldots, P [m..m]\) and suffixes starting at phrase boundaries.  This takes a total of $\Oh{m}$ time.  Finding the LCPs of \(P [1], (P [1..2])^\mathrm{rev}, \ldots, P^{\mathrm{rev}}\) is symmetric.

Suppose we already know the LCPs of \(P [1], (P [1..2])^\mathrm{rev}, \ldots, P^\mathrm{rev}\) and the reversed phrases, and the LCPs of \(P [2..m], \ldots, P [m]\) and the suffixes starting at phrase boundaries.  Then in a total of $\Oh{m \log^2 n}$ time we can find with high probability values $h$ and $j$ such that some phrase ends with \(P [h..i]\) and the next phrase starts with \(P [i + 1..j]\) and \(j - h + 1\) is maximum, for each choice of $i$.  To do this, we use $m$ applications of Theorem~\ref{thm:small HIAs} to $T_\mathrm{rev}$ and $T_\mathrm{suf}$, one for each partition of $P$ into a prefix and a suffix.  This gives us the following result:

\begin{theorem} \label{thm:medium LCSs}
Let $S$ be a string of length $N$ whose LZ77 parse consists of $n$ phrases.  We can store $S$ in $\Oh{n \log N}$ space such that, given a pattern $P$ of length $m$, we can find with high probability a longest substring common to $P$ and $S$ in $\Oh{m \log^2 n}$ time.
\end{theorem}

We can reduce the time bound in Theorem~\ref{thm:medium LCSs} to $\Oh{m \log n \log \log n}$ at the cost of increasing the space bound to $\Oh{n (\log N + \log^2 n)}$, by using the data structure from Theorem~\ref{thm:big HIAs} instead of the one from Theorem~\ref{thm:small HIAs}.  In fact, as we noted in Section~\ref{sec:HIAs}, in the full version of this paper we will also eliminate the \(\log \log n\) factor here.

\small
\bibliographystyle{abbrv}

\section*{Postscript} \label{sec:ps}

After submitting this paper, we realized how to use a new result by Bille et al.~\cite{BCGSVV13} to derandomize Ferragina's data structure~\cite{Fer13}, so we can now remove the words ``with high probability'' from Theorem~\ref{thm:medium LCSs}.  More importantly, we also realized that we can easily combine Theorem~\ref{thm:small HIAs} with the approach of Amir et al.~\cite{AKLLLR00} and Ferragina, Muthukrishnan and De Berg~\cite{FMD99} to obtain the following result; we will give more details in the full version of this paper.

\begin{theorem} \label{thm:error}
We can store a string $S$ of length $N$ in $\Oh{N}$ space such that later, given a pattern $P$ of length $m$, in $\Oh{m \log^{3 + \epsilon} N}$ time we can find a longest substring of $P$ that is within edit distance 1 of a substring of $S$.
\end{theorem}

\end{document}